\documentclass[letterpaper, 10 pt, conference]{ieeeconf}  

\def\compile{partA,partB,partC,partD,partE,partF,partG,partH,partI,partJ,partK,partL,partM,partN,partO,partP}    

\usepackage{xcomment}   

\newenvironment{partA}[1][]{}{}
\newenvironment{partB}[1][]{}{}  
\newenvironment{partC}[1][]{}{}
\newenvironment{partD}[1][]{}{}  
\newenvironment{partE}[1][]{}{}
\newenvironment{partF}[1][]{}{}
\newenvironment{partG}[1][]{}{}  
\newenvironment{partH}[1][]{}{}
\newenvironment{partI}[1][]{}{}  
\newenvironment{partJ}[1][]{}{}
\newenvironment{partK}[1][]{}{}  
\newenvironment{partL}[1][]{}{}
\newenvironment{partM}[1][]{}{}  
\newenvironment{partN}[1][]{}{}
\newenvironment{partO}[1][]{}{}
\newenvironment{partP}[1][]{}{}

\IEEEoverridecommandlockouts                              
                                                          
\usepackage[pdftex]{graphicx}
\usepackage[cmex10]{amsmath}
\usepackage{amsfonts}
\usepackage{color}
\usepackage{tikz}
\usetikzlibrary{automata}
\usepackage{comment}

\newtheorem{corollary}{Corollary}
\newtheorem{lemma}{Lemma}

\newtheorem{rem}{Remark}
\newtheorem{defin}{Definition}
\newtheorem{theor}{Theorem}
\newtheorem{exam}{Example}
\def\skp{\vskip0pt}

\newenvironment{remark}{\skp\vspace{-\lastskip}\par\addvspace{.6pc plus .2pc minus .1pc}\begin{rem}}{\hfill$\diamond$\end{rem}\par\addvspace{.6pc plus .2pc minus .1pc}\skp}
\newenvironment{definition}{\skp\vspace{-\lastskip}\par\addvspace{.6pc plus.2pc minus.1pc}\begin{defin}}{\hfill$\diamond$\end{defin}\par\addvspace{.6pc plus.2pc minus.1pc}\skp}
\newenvironment{theorem}{\skp\vspace{-\lastskip}\par\addvspace{.6pc plus.2pc minus .1pc}\begin{theor}}{\hfill$\square$\end{theor}\par\addvspace{.6pc plus .2pc minus .1pc}\skp}

\newcommand{\crr}[1]{\cr\noalign{\vskip#1pt}}
\newcommand{\R}{\mathbb{R}}
\newcommand{\N}{\mathbb{N}}
\renewcommand{\frac}[2]{{{\displaystyle#1}\over{\displaystyle#2}}}

\newenvironment{eqalign}[1]{\begin{equation}\begin{aligned}#1}{\end{aligned}\end{equation}}

\catcode`@=11
\def\eqalign#1{\null\,\vcenter{\openup\jot\m@th\ialign{\strut\hfil$\displaystyle{##}$&$\displaystyle{{}##}$\hfil\crcr#1\crcr}}\,}
\def\eqaligntwo#1{\null\,\vcenter{\openup\jot\m@th\ialign{\strut\hfil$\displaystyle{##}$&$\displaystyle{##}$\hskip.125cm&\hskip.125cm$\hfil\displaystyle{##}$&$\displaystyle{{}##}$\hfil\crcr#1\crcr}}\,}
\def\eqalignthree#1{\null\,\vcenter{\openup\jot\m@th\ialign{\strut\hfil$\displaystyle{##}$&$\displaystyle{##}$\hskip.25cm&\hskip.25cm$\hfil\displaystyle{##}$&$\displaystyle{{}##}$\hskip.25cm&\hskip.25cm$\hfil\displaystyle{##}$&$\displaystyle{{}##}$\hfil\crcr#1\crcr}}\,}
\def\eqalignno#1{\displ@y \centering\halign to\displaywidth{\hfil$\@lign\displaystyle{##}$&$\@lign\displaystyle{{}##}$\hfil\centering&\llap{$\@lign##$}\crcr#1\crcr}}
\def\tabletwo#1{\null\,\vcenter{\openup\jot\m@th\ialign{\strut\hfil$\displaystyle{##}$&$\displaystyle{##}$\hskip.125cm&\hskip.125cm$\hfil\displaystyle{##}$&$\displaystyle{{}##}$\hfil\crcr#1\crcr}}\,}
\def\matrix#1{\null\,\vcenter{\normalbaselines\m@th
    \ialign{\hfil$##$\hfil&&$\>\>$\hfil$##$\hfil\crcr
      \mathstrut\crcr\noalign{\kern-\baselineskip}
      #1\crcr\mathstrut\crcr\noalign{\kern-\baselineskip}}}\,}
\def\pmatrix#1{\left(\matrix{#1}\right)}

\catcode`@=12

\begin{document}
                                                                \begin{xcomment}{\compile}
                                                                \begin{partA}


\title{\LARGE\bf Event--Triggered Observers and Observer--Based Controllers\\for a Class of Nonlinear Systems}

\author{L. Etienne, S. Di Gennaro, and J.--P. Barbot\thanks{Lucien Etienne and S. Di Gennaro are with the Department of Information Engineering, Computer Science and Mathematics, and with the Center of Excellence DEWS, University of L{'}Aquila, Via Vetoio, Loc. Coppito, 67100 L{'}Aquila, Italy. E.mail: {\tt \{lucien.etienne, stefano.digennaro\}@univaq.it.}}
\thanks{Jean--Pierre Barbot is with the Ecole Nationale Superieure d'Electronique et de Ses Applications -- ENSEA, 6 Avenue du Pon\c ceau, 95014 Cergy Pontoise Cedex E.mail: {\tt barbot@ensea.fr}.}
}

\maketitle

\begin{abstract}

In this paper, we investigate the stabilization of a nonlinear plant subject to network constraints, under the assumption of partial knowledge of the plant state. The event triggered paradigm is used for the observation and the control of the system. Necessary conditions, making use of the ISS property, are given to guarantee the existence of a triggering mechanism, leading to asymptotic convergence of the observer and system states. The proposed triggering mechanism is illustrated in the stabilization of a robot with a flexible link robot.
\end{abstract}

\section{Introduction}

The use of the digital technology is pervasive in modern control systems, where the control task consists of the sampling of the plant outputs, the computation, and the implementation of the actuator signals. The classic way is to sample in a periodic fashion, thus allowing the closed--loop system to be analysed on the basis of sampled--data systems, see~\cite{Astrom 1997}. Recent years have seen the development of a different paradigm where, instead of being sampled periodically (i.e. with a time--triggered policy), the system is triggered when the stability property is lost (i.e with an event--triggered policy). A good number of works deal with this subject, see \cite{Astrom 2003}, \cite{Tabuada 2007}, \cite{Wang 2011}, \cite{Lunze 2010}, \cite{Heemels 2008}, and \cite{Heemels 2012} for an introduction to this topic. The problem is to design an event--triggered mechanism to ensure the closed--loop stability. This problem was solved, for both the linear and the nonlinear case, when the full state is available~\cite{Tabuada 2007}, \cite{Wang 2011}. When the  state is not available, the problem was addressed in~\cite{Lehmann 2011b}, \cite{Donkers 2012} for linear systems. In~\cite{Tallapragada 2013} the results were extended to linear event--triggered network control systems. In the nonlinear setting, to the best of the authors' knowledge, no result is still available when the whole state is not available for feedback.

The main objective of this paper is to address the problem of the event--triggered output--based feedback for nonlinear systems, giving sufficient conditions for the dynamic feedback control of nonlinear plants subject to network constraints, using an event--triggered strategy.

The paper is organized as follows. In Section~II we recall the event--triggered control, and we introduce the class of systems considered. In Section~III we give sufficient conditions on the observer and on the observation error in terms of input--to--state stability, along with relevant event--triggering mechanisms, in order to ensure asymptotic convergence to the origin. In Section~IV we consider some type of systems fitting into the class of systems considered in Section~III. In Section~V an example is given. Finally, in Section~VI we give some concluding remarks.

\medskip
\noindent{\sl Notation:} In the following, $|\cdot|$ denotes the norm $\|\cdot\|_1$, and $\|\cdot \|$ is the euclidean norm. Moreover, $|\cdot|_{\infty}$ is the component with the biggest absolute value. Furthermore, $\alpha(\cdot)\in {\cal K}$  if it a strictly increasing function from $[0,a) \to [0,\infty)$, while $\alpha$ is a class ${\cal K}_{\infty}$ function if it  strictly increasing function from $[0,\infty)\to [0,\infty)$ and $\lim_{r\to \infty} \alpha(r)=\infty$. Finally, $\beta(r,s)\in {\cal KL}$ if $\beta(\cdot,s)\in {\cal K}$ for all $s$ and $\lim_{s\to\infty}\beta(r,s)=0$ for all $r$. When a function $f$ is Lipschitz, we denote $L_f$ its Lipschitz constant.

\section{Problem formulation and definitions}

                                                                \end{partA}
%
%
%
                                                                \begin{partB}

\subsection{Problem Statement and Event Triggering Policies}

                                                                \end{partB}
%
%
%
                                                                \begin{partC}

We will first recall some known facts and terminologies about event triggered systems. Consider the system 
\begin{equation}
\eqalign{
\dot x(t)&=f_s(x(t),u(t)) \crr{0}
y(t)&= h(x(t))   \cr}                           \label{nonlinear system}
\end{equation}
where $x\in\R^n$ is the state, $u\in\R^m$ is the control, $y\in\R^p$ is the output. The time instant $t$ is dropped if there are no ambiguities. The functions $f$ and $h$ are assumed sufficiently smooth. We also assume the existence of a continuous state based controller which renders the origin asymptotically stable.

The control scheme is shown in Fig.~\ref{feedback scheme}. Due to the \emph{communication constraints}, there is no continuous communication either between sensors and observer, or between observer and actuators. The inputs and the outputs are partitioned into actuator/sensor nodes $u=(u_1^T,\cdots,u_q^T)^T$, $y=(y_1^T,\cdots,y_r^T)^T=(h_1^T(x),\cdots,h_r^T(x))^T$, with $u_1,\cdots,u_q$, $y_1,\cdots, y_r$, not necessarily scalars.

The value $y_i(t_{k_i})=h_i(x(t_{k_i}))$, $i=1,\cdots,r$, is the last sampled value at the $i^{th}$ sensor node, available for the controller to implement the control, while the value $u_i(t_{j_i})$, $i=1,\cdots,q$, is applied to the system at the $i^{th}$ actuator node, through a classic zero--order holder $H_0$. It is worth noting that this means that the different outputs $\{y_i\}_{i=1,\cdots,r}$ and the different inputs $\{u_i\}_{i=1,\cdots,q}$ are not sampled synchronously. For this reason, at time $t$ the latest output available is 
$$
\bar y(t)=\Big(y_1^T(t_{k_1}),y_2^T(t_{k_2})\cdots,y_p^T(t_{k_p})\Big)^T
$$
while the control is 
$$
\bar u(t)=\Big(u_1^T(t_{j_1}),u_2^T(t_{j_2})\cdots, u_q^T(t_{j_q})\Big)^T.
$$
Denoting by $e_u=u-\bar u$ and $e_y=y-\bar y$ the difference vectors between the continuous and sampled values, one considers the vector $E=(e_u^T,e_y^T)^T$ of the error due to the sampling.

\begin{figure}[h!]
\centerline{\includegraphics[width=8.7cm]{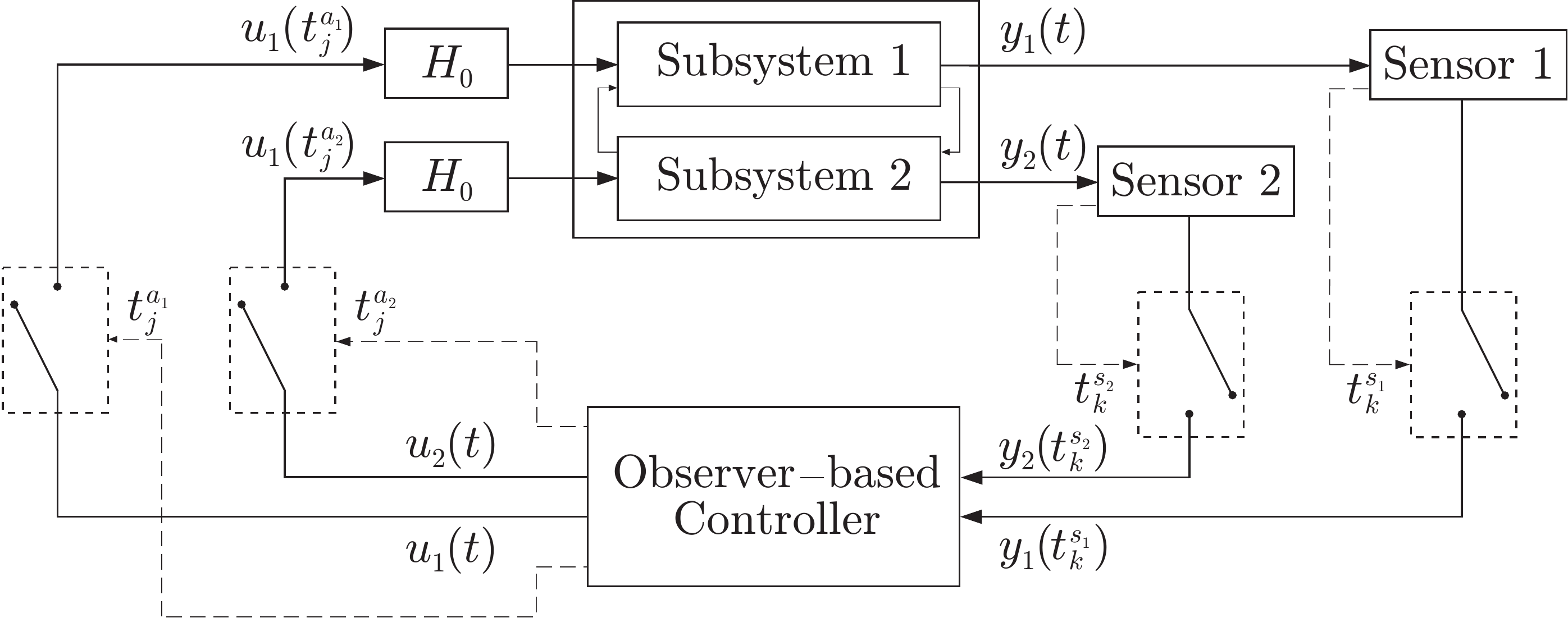}} 
\caption{Control scheme with sampled output and zero order holder}
\label{feedback scheme}
\end{figure}

Let us consider first a simple case in which the state $x$ is available for measurement, and let us assume that there exists a state--feedback 
\begin{equation}
u=\gamma(x)  \label{ctrlc}
\end{equation}
rendering system~\eqref{nonlinear system} asymptotically stable at the origin. The partitioned input vector is 
$$
u=\Big(\gamma_1^T(x),\cdots,\gamma_q^T(x)\Big)^T. 
$$
When the controller is implemented making use of the sampled values, one considers the last communication time $t_j$ between controller and plant, and the control value
$$
\bar u=\bar\gamma(x)=\Big(\gamma_1^T(x(t_{j_1}),\cdots,\gamma_q^T(x(t_{j_q}))\Big)^T.
$$

Using a classic periodic sampling, the next sampling time is $t_{k+1}=t_k+\delta$, where $\delta>0$, so that $t_{k+1}-t_k=\delta>0$ or, that is the same
$$
t_{k+1}=\min_t \{t\mid t>t_k+\delta\}. 
$$
The event triggered paradigm replaces this condition with a condition on the state values $x(t), x(t_k)$. A simple condition of this kind is, for instance, the epsilon crossing policy, which is of the form
$$
t_{k+1}=\min_t \{t\ge t_k\mid |x(t)-x(t_k)|>\varepsilon\}
$$
viz. $x(t)$ is sampled when $|x(t)-x(t_k)|$ is greater than a certain threshold value $\varepsilon\in\R$. When this condition is verified, an event is triggered, which determines the sampling time $t_{k+1}$. The difference $\delta_k=t_{k+1}-t_k$ is usually called the inter--event time. To avoid Zeno behaviors~\cite{Johansson 1999}, it is important that the chosen sampling policy ensures that $\delta_k>0$ for all $k\in\N$, possibly under additional conditions.

Further strategies can also be used to determine the next sampling time. For instance, the state dependent triggering condition
$$
t_{k+1}=\min_t \{t\ge t_k\mid|x(t)-x(t_k)|>\sigma|x|+\varepsilon\}
$$
with $\varepsilon,\sigma\in\R^+$, or a mixed triggering policy 
$$
t_{k+1}=\min_t \{t\ge t_k+\delta_{\min},\mid |x(t)-x(t_k)|>\varepsilon\}
$$
with $\varepsilon,\delta_{\min}\in\R^+$. Furthermore, \eqref{nonlinear system} can be stabilized asymptotically with the state triggering condition 
$$
t_{k+1}=\min_t \{t\ge t_k,\mid |x(t)-x(t_k)|>\sigma |x(t)|\}
$$ 
under the sole assumption that the closed loop nonlinear system is input--to--state stable with respect to the quantity $|x(t)-x(t_k)|$~\cite{Tabuada 2007}.

When the state $x$ of~\eqref{nonlinear system} is not measurable, these triggering policies cannot be implemented. In the following, we will introduce the triggering policy that will be used in this case, taking into account the constraints on the communication of output and input. An obvious assumption is that it is possible to design an observer that converges asymptotically to $x$, of the form  
$$
\dot{\hat x}=f_o(\hat x,y,u).
$$
where $f_o:\R^n\times\R^p\times\R^m\to\R^n$ is not smooth, in general. In view of an implementation via a triggering policy, and since the observer has not $y(t)$ available, one can use the vector $\bar y$, so considering the observer
\begin{equation}
\dot{\hat x}=f_o(\hat x,u,\bar y).   \label{observer}
\end{equation}

A feedback controller based on $\hat x$ given by~\eqref{observer} will be used in the following to stabilize the system~\eqref{nonlinear system} in the origin. The input applied to the system, due to the communication channel, is $\bar u=\bar\gamma(\hat x)$, so obtaining the controlled dynamics
$$
\dot x= f_s(x,\bar\gamma(\hat x)).
$$
Eventually, one gets the following closed--loop system
$$
\begin{aligned}
\dot x&= f_s(x,\bar\gamma(\hat x))                      \\
\dot{\hat x}&= f_o(\hat x,\bar\gamma(\hat x),\bar y).    
\end{aligned}
$$

The observation error is $z=x-\hat x$. We assume that the observation error dynamics can be written is the form
$$
\dot z=f_s(x,\bar\gamma(\hat x))- f_o(\hat x,\bar\gamma(\hat x),\bar y)=g(z,\theta_1(e_u),\theta_2(e_y),\hat x)
$$
where $\theta_1,\theta_2$ give the dependence on the input and the output errors $e_u$, $e_y$, due to the sampling.

                                                                \end{partC}
%
%
%
                                                                \begin{partD}


\subsection{Definitions}

\begin{definition}[Input--to--state stability--ISS]
System~\eqref{nonlinear system} is said to be locally ISS if there exist a ${\cal KL}$ function $\beta$, a ${\cal K}$ function $\alpha$, and some constants $k_1,k_2>0$
such that 
$$ 
|x(t)|\le\beta(|x_0|,t)+\alpha(|u|),\quad \forall\, t\ge0
$$ 
for all $x_0\in D$, $u\in D_u$ satisfying $|x_0|<k_1$, $|u|<k_2$. System~\eqref{nonlinear system} is said (globally) input--to--state stable if $D=\R^n$, $D_u=\R^m$, and the above inequalities are satisfied for any initial state and any bounded input.
\end{definition}
\medskip


\medskip
\begin{definition}[ISS Lyapunov function]
 $\text{A continuous function}$ $V:D\to R$ is an ISS Lyapunov function on $D$ for system~\eqref{nonlinear system} if there exist class
${\cal K}$ functions $\alpha_1,\alpha_2,\alpha_3,\beta$ such that the following two conditions are satisfied
$$
\begin{aligned}
\alpha_1(|x|)&\le V(x(t))\le \alpha_2(|x|)     & &\forall\, x \in D, t\ge0   \\
\dfrac{\partial V(x)}{\partial x}f(x, u)&\le -\alpha_3(|x|) +\beta(|u|) & &\forall\, x\in D, u \in D_u.
\end{aligned}
$$
Moreover, $V$ is an global ISS Lyapunov function if $D=R^n,D_u=R^m$, and $\alpha_1,\alpha_2,\alpha_3, \beta, \in {\cal K}_{\infty}$.
\end{definition}

                                                                \end{partD}
%
%
%
                                                                \begin{partE}

\section{Main result}

\subsection{Hypothesis on the Dynamics of the State Observer and of the Observation Error}

Since the observer state is available, in the following we consider the observer dynamics, so allowing imposing on $\hat x$ a triggering condition, along with the observation error dynamics
\begin{subequations}\label{closed-loop system}
\begin{align}
\dot{\hat x}&= f_o(\hat x,\bar\gamma({\hat x}),\bar h(\hat x+z))  \label{closed-loop system:1}\\
\dot z&=g(z,\theta_1(e_u),\theta_2(e_y),\hat x)                    \label{closed-loop system:2}
\end{align}
\end{subequations}
where $y=h(\hat x+z)$ and $\bar y=\bar h(\hat x+z)$, or equivalently
\begin{equation}
\dot X=G(X,E)     \label{extended dynamics}
\end{equation}
where $X=(\hat x^T,z^T)^T$ is an extended state vector, and $G=(f_o^T,g^T)^T$. In the following we consider the following assumptions.

\medskip
\begin{enumerate}
\item[$(A_1)$] There exists an ISS Lyapunov function for~\eqref{closed-loop system:1} such that $\forall\, \hat x,z\in\R^n,E \in\R^{m+p}$, $\forall\, t\ge0$
$$
\begin{aligned}
\alpha_{c,1}(|\hat x|)&\le V_c(\hat x(t))\le \alpha_{c,2}(|\hat x|) \\
\dfrac{\partial V_c(\hat x)}{\partial \hat x} f_o(&\hat x,\bar\gamma(\hat x),\bar h(\hat x+z))\le -\alpha_{c,3}(|\hat x|)+\beta_c(|(z,E)|)
\end{aligned}
$$ 
with $\alpha_{c,1},\alpha_{c,2},\alpha_{c,3},\beta_c\in{\cal K}$, and $\beta_c,\alpha_{c,3}^{-1}$ Lipschitz;

\item[$(A_2)$] There is an ISS Lyapunov function for \eqref{closed-loop system:2} such that $\forall\,z \in\R^n,E \in\R^{m+p}$, $\forall\, t\ge0$
$$
\begin{aligned}
\alpha_{o,1}(|z|)&\le V_o(z(t))\le \alpha_{o,2}(|z|)   \\
\dfrac{\partial V_o(z)}{\partial z}g(&z,\theta_1(e_u),\theta_2(e_y),\hat x)\le -\alpha_{o,3}(|z|) +\beta_o(|E|)
\end{aligned}
$$
with $\alpha_{o,1},\alpha_{o,2},\alpha_{o,3},\beta_o\in{\cal K}$, and $\beta_o,\alpha_{o,3}^{-1}$ Lipschitz;

\item[$(A_3)$] $f_o$, $h$ and $\gamma$ are Lipschitz;

\item[$(A_4)$] $g$ is Lipschitz with respect to $(z,\theta_1(e_u),\theta_2(e_y))$, uniformly in $\hat x$, and $\theta_1,\theta_2$ are Lipschitz.
\end{enumerate}

\medskip
\begin{remark}
$(A_1)$ ensures the asymptotic convergence to the origin of the observer, in absence of sampling errors and observation error, and an ISS property with respect to $z,e_u,e_y$. $(A_2)$ ensures the asymptotic convergence to zero of the observation error in absence of sampling errors, and an ISS property with respect to $e_u,e_y$.
Those two assumption suppose a separation principle between state estimation and control.
\end{remark}
\medskip

Since we are interested in the stabilisation of the observer state $\hat x$ and of the observation error state $z$, in the following we will assume that $X(0)\ne 0$.

                                                                \end{partE}
%
%
%
                                                                \begin{partF}

\medskip
\begin{lemma}\label{lemma1}  
$\text{Under the Assumptions}$~$(A_1),(A_2),(A_3),(A_4)$, the extended system $X=(\hat x^T,z^T)^T$ admits an ISS Lyapunov function $V(X)$ such that $\forall\, X \R^{2n},\forall E\in R^{n+p}$, $\forall\,t\ge0$
$$
\begin{aligned}
a_1(|X|)&\le V(X)\le a_2(|X|)  \\
\dfrac{\partial V(X)}{\partial X}G(X,E)&\le -a_3(|X|) +b(|E|) 
\end{aligned}
$$
with $a_1,a_2,a_3,b\in{\cal K}$, $b,a_3^{-1},G$ Lipschitz.
\end{lemma}

                                                                \end{partF}
%
%
%
                                                                \begin{partG} 

\medskip
\begin{proof}
Let us consider the candidate ISS Lyapunov function 
$$
V(X)=\lambda_c V_c(\hat x)+V_o(z).
$$
From $(A_1),(A_2)$, 
$$
\begin{aligned}
a_1(|X|)&=\min_{|(\hat x,z)|=|X|} \lambda_c \alpha_{c,1}(|\hat x|)+\alpha_{o,1}(|z|)\\
        &\le \lambda_c \alpha_{c,1}(|\hat x|)+\alpha_{o,1}(|z|)                     \\
a_2(|X|)&=\max_{|(\hat x^T,z^T)|=|X|} \lambda_c \alpha_{c,2}(|\hat x|)+\alpha_{o,2}(|z|)\\
        &\ge \lambda_c \alpha_{c,2}(|\hat x|)+\alpha_{o,2}(|z|)
\end{aligned}
$$
with $a_1,a_2\in{\cal K}$. Furthermore,
$$
\begin{aligned}
\dfrac{\partial V(X)}{\partial X}G(X,E)&=\left(\dfrac{\partial V(X)}{\partial \hat x}\ \dfrac{\partial V(X)}{\partial z}\right)\pmatrix{f_o\crr{2} g}  \\
           &\le \lambda_c\Big(-\alpha_{c,3}(|\hat x|)+\beta_c(|(z^T,E^T)^T|) \Big)\\
           &\qquad  -\alpha_{o,3}(|z|) +\beta_o(|E^T|) \\
           &\le -\Big(\lambda_c \alpha_{c,3}(|\hat x|)+\alpha_{o,3}(|z|)-\lambda_c L_{\beta_c}|z|\Big)\\
           &\qquad +\lambda_c L_{\beta_c}|E|+\beta_o(|E|)
\end{aligned}
$$
where we have used the fact that
$$
\beta_c(|(z^T,E^T)^T|)  \le L_{\beta_c}|(z^T,E^T)^T| \le L_{\beta_c}|E|+L_{\beta_c}|z|.
$$
It is always possible to choose $\lambda_c$ sufficiently small such that $\alpha_{o,3}(|z|)-\lambda_c L_{\beta_c}|z|$ is a class ${\cal K}$ function with $z$ as variable. Since we are considering 1--norm
$$
a_3(|X|)=\min_{|X|}\Big\{\lambda_c\alpha_{c,3}(|X|),\alpha_{o,3}(X)-\lambda_c L_{\beta_c}|X|\Big\}.
$$
To show that $a_3^{-1}$ is Lipschitz, note first that since $\alpha_{o,3}^{-1}$ is Lipschitz
$$
\alpha_{o,3}(|z|)\ge \frac{1}{L_{\alpha_{o,3}^{-1}}}|z|.
$$
Moreover, one can compute an upper bound on the derivative of $(\alpha_{o,3}(|\cdot|)-\lambda_c L_{\beta_c}|\cdot|)^{-1}$, since 
$$
\frac{d}{d|z|}(\alpha_{o,3}(|z|)-\lambda_c L_{\beta_c}|z|)^{-1}\le \dfrac{L_{\alpha_{o,3}^{-1}}}{1-\lambda_c L_{\alpha_{o,3}^{-1}} L_{\beta_c}} 
$$
Hence it is always possible to choose $\lambda_c$ sufficiently small such that $(\alpha_{o,3}(|\cdot|)-\lambda_c L_{\beta_c}|\cdot|)^{-1}$ is a class ${\cal K}$ function with Lipschitz constant 
$$
L_{a_3^{-1}}=\max\left\{\frac{\lambda_c}{L_{\alpha_{c,3}}},\dfrac{L_{\alpha_{o,3}^{-1}}}{1-\lambda_c L_{\alpha_{o,3}^{-1}} L_{\beta_c}}\right\}.
$$
Furthermore,
$$
b(|E|)= L_{\beta_c}|E|+\beta_o(|E|)
$$ 
which is Lipschitz with constant $L_b=L_{\beta_c}+L_{\beta_o}$. Finally, thanks to $(A_3),(A_4)$, $G(X,E)$ is Lipschitz.
\end{proof}

                                                                \end{partG}
%
%
%
                                                                \begin{partH}

\medskip
In the following section we are interested in providing sufficient conditions on the stabilisation of a nonlinear system using the event trigger paradigm. The key concept will be the ISS of both the closed--loop system and of the observer dynamics. For, we introduce the following lemmas.

\medskip
\begin{lemma}\label{lemma2}
If the observer and the error dynamics verify $(A_1),(A_2),(A_3),(A_4)$, then there exist a $\sigma>0$ such that any sampling policy ensuring $|E|\le \sigma |X|$, leads to asymptotic convergence of the overall system to the origin.
\end{lemma}

                                                                \end{partH}
%
%
%
                                                                \begin{partI} 

\medskip
\begin{proof} From Lemma~\ref{lemma1}, the existence of an ISS Lyapunov function $V$ is ensured. Since $a_3^{-1}$ and $b$ are Lipschitz, $a_3(|X|)\ge \dfrac{1}{L_{a_3^{-1}}}|X|$
and 
$$
-a_3(|X|)+b(\sigma|X|) \le -\left(\frac{1}{L_{a_3^{-1}}}-L_b\sigma\right)|X|.
$$
Therefore  for all $\sigma \in \left(0 ,\dfrac{1}{L_{a_3^{-1}}L_b}\right)$ the system~\eqref{closed-loop system} converge asymptotically to the origin.
\end{proof}

                                                                \end{partI}
%
%
%
                                                                \begin{partJ}

\medskip
\begin{remark}
Under the hypothesis that $a_2,a_1^{-1}$ are Lipschitz, one can prove exponential convergence of \eqref{closed-loop system}. In fact, since 
$$
\frac{1}{L_{a_1^{-1}}}|X|< a_1(|X|)<V(|X|)
$$
one has that 
$$
\dot V(|X|)\le -\left(\frac{1}{L_{a_3^{-1}}}-L_b\sigma\right)|X|.
$$
Therefore
$$
\dot V(|X|)\le-\left(\frac{1}{L_{a_3^{-1}}}-L_b\sigma\right)L_{a_1^{-1}}V(|X|).
$$
\end{remark}

\medskip
\begin{remark}
The choice $\sigma \in\left(0 ,\dfrac{1}{L_{a_3^{-1}}L_b}\right)$ represents a trade--off between the sampling rate and the convergence rate.
\end{remark}

Since $|E|<\sigma |X|$, using the norm equivalence there exists a $\sigma'>0$ such that $\|E\|<\sigma'\|X\|$ implies $|E|<\sigma |X|$.

\medskip
\begin{lemma}\label{lemma3} For every $\kappa_i>0$ there is a minimal time $\tau_{\min}>0$ such that if $|E|\le \sigma|X|$, then $\forall \, t_k$, $\forall \, t\in [t_k,t_k+\tau_{\min})$ the following inequalities are verified
$$
\begin{aligned}
\|\gamma_i(\hat x(t))-\gamma_i(\hat x(t_k))\|&\le \kappa_i\|X\|  \\
\|h_j(\hat x(t)+z(t))-h_j(\hat x(t_k)+z(t_k)\|&\le\kappa_j\|X\|.
\end{aligned}
$$
\end{lemma}

                                                                \end{partJ}
%
%
%
                                                                \begin{partK} 

\medskip
\begin{proof} In the following we assume $X\ne 0$. The argument follows the proof of Theorem~1 in \cite{Tabuada 2007}. Denoting $e_{u_i}=\gamma_i(\hat x(t))-\gamma_i(\hat x(t_k))$, one works out
$$
\begin{aligned}
\dfrac{d}{dt}\dfrac{\|e_{u_i}\|}{\|X\|}&=\dfrac{e_{u_i}^T\dot e_{u_i}}{\|e_{u_i}\| \|X\|}-\dfrac{X^T\dot X\|e_i\|}{\|X\|^3} \\
                                       &\le \dfrac{\|e_{u_i}\|\|\dot e_{u_i}\|}{\|e_{u_i}\| \|X\|}+\dfrac{\|\dot X\|\|e_i\|}{\|X\|^2}.
\end{aligned}                                       
$$
Since $\|\dot e_{u_i}\|\le L_{\gamma_i} \|\dot{\hat x}\|\le L_{\gamma_i} \|\dot X\|$,
$$
\dfrac{d}{dt}\dfrac{\|e_{u_i}\|}{\|X\|}\le \dfrac{\|\dot{X}\|}{\|X\|}\left(L_{\gamma_i}+\dfrac{\|e_i\|}{\|X\|}\right)
$$
Moreover, $G$ is Lipschitz, so that
$$
\dfrac{d}{dt}\dfrac{\|e_{u_i}\|}{\|X\|}\le \dfrac{L_G(\|X\|+\|E\|)}{\|X\|}\left(L_{\gamma_i}+\dfrac{\|e_i\|}{\|X\|}\right)
$$
Since $\|E\|< \sigma' \|X\|$,
$$
\dfrac{d}{dt}\dfrac{\|e_{u_i}\|}{\|X\|}\le L_G(1+\sigma')\left(L_{\gamma_i}+\dfrac{\|e_i\|}{\|X\|}\right).
$$
At each reset time one has $e_{u_i}=0$. Using the comparison lemma with the differential equation 
$$
\dot y=L_G(1+\sigma')\big(L_{\gamma_i}+y\big),\quad y(0)=0
$$
one has
$$
\dfrac{\|e_{u_i}(t)\|}{\|X(t)\|}\le \big(e^{L_G(1+\sigma'(t-t_k)}-1\big)L_{\gamma_i} .
$$
Therefore the inequality 
$$
\|\gamma_i(\hat x(t))-\gamma_i(\hat x(t_k))\|>\kappa_i\|X\|
$$
can not be true before time 
$$
\tau_{\min}^i=\dfrac{1}{L_G(1+\sigma')}\ln\big(1+\dfrac{\kappa_i}{L_{\gamma_i}}\big).
$$
Analogously, for 
$$
\|h_j(\hat x(t)+z(t))-h_j(\hat x(t_k)+z(t_k)\|>\kappa_j\|X\|
$$
gives for the sensors 
$$
\tau_{\min}^j=\dfrac{1}{L_G(1+\sigma')}\ln\big(1+\dfrac{\kappa_j}{L_{h_i}}\big).
$$
\end{proof}

                                                                \end{partK}
%
%
%
                                                                \begin{partL}

\medskip
Let us define the triggering function at each node
\begin{equation}
t_{k+1}^i=\min_t \{t\ge t_k^i+\tau_{\min}^i,\mid  \|u_i(t)-u_i(t_k^i)\|>\kappa_i\|X\|\}   \label{ctrl}
\end{equation}
\begin{equation}
t_{k+1}^j=\min_t \{t\ge t_k^j+\tau_{\min}^j,\mid \|y_j(t)-y_j(t_k^i)\|>\kappa_j\|X\|\}.    \label{obs}
\end{equation}

\medskip
\begin{remark} From Lemma~\ref{lemma3}, $t_{k+1}^i=\min_t \{t\ge t_k^i+\tau_{\min}^i,\mid  \|u_i(t)-u_i(t_k^i)\|>\kappa_i\|X\|\}=t_{k+1}^i=\min_t \{t\ge t_k^i,\mid \|u_i(t)-u_i(t_k^i)\|>\kappa_i\|X\|\}$.
\end{remark}

\medskip
\begin{lemma}
If $\sum_{\{1,\cdots,r\}\cup\{1,\cdots,q\}}\kappa_i\le \sigma'$ then \eqref{ctrl} and \eqref{obs} ensure $\|E\|\le \sigma\|X\|$.
\end{lemma}

                                                                \end{partL}
%
%
%
                                                                \begin{partM}

\medskip
\begin{proof} from \eqref{ctrl} and \eqref{obs}
$$
\|E\|\le \sum_{\{1,\cdots,r\}\cup\{1,\cdots,q\}}\kappa_i\|X\|\le \sigma'\|X\|.
$$
\end{proof}

                                                                \end{partM}
%
%
%
                                                                \begin{partN}

\medskip
The proposed triggering conditions allow asymptotic convergence with a nonzero minimum inter--event time. Unfortunately, they are not implementable on a network for two reasons. The first is that $X$ is not available, since the observation error is not known. The second is that sensors do not communicate among them nor receive information from the observer--based controller. Nevertheless, considering the following modified triggering conditions
\begin{equation}
t_{k+1}^i=\min_t \{t\ge t_k^i+\tau_{\min}^i,\mid  \|u_i(t)-u_i(t_k^i)\|>\dfrac{\kappa_i}{L_{\gamma_i}}\|\gamma_i(\hat x)\|\}\label{ctrlp}
\end{equation}
\begin{equation}
t_{k+1}^j=\min_t \{t\ge t_k^j+\tau_{\min}^j,\mid  \|y_j(t)-y_j(t_k^i)\|>\dfrac{\kappa_j}{2L_h}\|y_j\|\}\label{obsp}
\end{equation}
this approach can be used on a network, allowing asymptotic convergence and a nonzero minimal inter--event time, using only information available at each node, as stated by the main contribution of this work.

\medskip
\begin{theorem}\label{theorem1}
If $(A_1),(A_2),(A_3),(A_4)$ are verified, and the sampling instants are defined by \eqref{ctrlp}, \eqref{obsp}, then the origin of the closed--loop system~\eqref{closed-loop system} is asymptotically stable and there exists a nonzero minimum inter--event time for each node.
\end{theorem}

\medskip
\begin{proof} Under the hypotheses of the theorem, Lemma~\ref{lemma1} applies. Since $\dfrac{\|y_j\|}{2L_h}< \|X\|$
and $ \dfrac{\|u_i\|}{L_{\gamma_i}}< \|X\|$ and  $\forall i \in\{1,\cdots,r\}$ from Lemma~\ref{lemma3} one can state that between $t_{k_i}$ and $t_{k_i}+\tau_{\min}^i$, $\|u_i(t)-u_i(t_k^i)\|>\kappa_i\|X\|$, while $\forall j \in\{1,\cdots,q\}$ between $t_{k_j}$ and $t_{k_j}+\tau_{\min}^j$, one has $\|y_j(t)-y_j(t_{k_j})\|>\kappa_j\|X\|$.

Therefore, $\|E\|<\sigma'\|X\|$. Using Lemma~\ref{lemma2}, there is asymptotic convergence of~\eqref{closed-loop system} to the origin.
\end{proof}

\section{Examples of Systems Fitting into the Proposed Framework}

\subsection{Linear Systems}

Let us consider a detectable and stabilizable linear system
\begin{equation}
\eqalign{
\dot{x}&=Ax+Bu \crr{0}
y&=x  \cr}                           \label{linear system}
\end{equation}
with 
\begin{equation}
\dot{\hat x}=A\hat x+Bu +LC(\bar x-\hat x) \label{linear system}
\end{equation}
a Luenberger observer. With control $K\bar{\hat x}$, one gets
$$
\begin{aligned}
\dot{\hat x}&=(A+BK)\hat x +BK(\bar{\hat x}-\hat x)+ LCz -LC(x-\bar{x})    \\
\dot z&=(A-LC)z +LC(x-\bar{x}).
\end{aligned}
$$
Since $A+BK$ and $A-LC$ are Hurwitz, it is possible to find an ISS Lyapunov function for the extended system. 


\subsection{Nonlinear Lipschitz Systems}

Let us consider a nonlinear Lipschitz system 
\begin{equation}
\begin{aligned}
\dot x&=Ax+Bu+\phi(x,u) \\
y&=Cx.
\end{aligned} \label{eqlip}                   
\end{equation}
Several results are available for the observer synthesis of nonlinear Lipschitz systems when the control and the output are implemented in a continuous fashion. We consider an observer of the form
\begin{equation}
\dot{\hat x}=Ax +BK\hat x+\phi(x,K\hat x).   \label{closed-loop nonlinear system:1}
\end{equation}
Hence, the extended closed--loop system is
\begin{subequations}
\label{closed-loop  nonlinear system}
\begin{align}
\dot{\hat x}&=A+\hat x +BK\hat x+\phi(\hat x,K\hat x)+LCz    \label{closed-loop nonlinear system:2} \\
\dot z&=(A-LC)z +\phi(x,K\hat x)-\phi(\hat x,K\hat x).        \label{closed-loop nonlinear system:3}
\end{align}
\end{subequations}
To implement an event--triggered control strategy, we need to consider the following structural properties.
\begin{enumerate}
\item[$(H_1)$] $\|\phi(x_1, u)-\phi(x_2, u)\| \le  \rho \|x_1-x_2\|$ , $\forall u \in \R^p$, $(x_1,x_2) \in \R^{2n}$;
\item[$(H_2)$] $\|\phi(x, u)\| \le  \rho \|x\| , \forall u \in\R^p$;
\item[$(H_3)$] There exist a gain $K$ such that $u=Kx$ for the system \eqref{closed-loop nonlinear system:1} and there exist a quadratic Lyapunov function
\begin{equation}
V_c(x)=x^TP_cx ,\quad\dot V_c(x)\le -\eta_c x^Tx   \label{Vc}
\end{equation} 
with $P_c=P_c^T>0$, $\eta_c>0$;
\item[$(H_4)$] There exist gain $L$ such that for \eqref{closed-loop nonlinear system:3} and there exist  quadratic Lyapunov function for the $z$ dynamic 
\begin{equation}
V_o(z)=z^TP_oz, \quad \dot V_o(z)\le-\eta_o z^Tz        \label{Vo} 
\end{equation}
with $P_o=P_o^T>0$, $\eta_o>0$.
\end{enumerate}

In $(H_2)$, for $\rho=0$ we have a linear system, and the existence of $V_c,V_o$ derive from the stabilizability and the detectability. Moreover, there always exists a $\rho_{\max}>0$ small enough such that the proposed Lyapunov function exist forall $\rho \in [0,\rho_{\max}]$. For other (more complex) conditions of existence of $V_c,V_o$ verifying \eqref{Vc}, \eqref{Vo}, see for instance~\cite{Pagilla 2004}.

\medskip 
\begin{lemma}\label{lemma5}
If $(H_1),(H_2),(H_3),(H_4)$ are verified, then the proposed observer and the observation error verify $(A_1),(A_2),(A_3),(A_4)$ .
 \end{lemma}

                                                                \end{partN}
%
%
%
                                                                \begin{partO}

\medskip
\begin{proof}
When subject to the trigger conditions, the observer has the following dynamics
$$
\dot{\hat x}= (A+BK)\hat x+BK(\hat x-\bar{\hat x})+LCz+LC(\bar{z}-z).
$$
Let us consider the candidate ISS Lyapunov function
$$
2\sqrt{\lambda_{\min}(P_c)}\|\hat x\|\le 2\sqrt{ V_c(\hat x)}\le2 \sqrt{\lambda_{\max}(P_c)}\|\hat x\|
$$
having derivative
$$
\eqalign{
&\frac{d}{dt} 2\sqrt{\hat x^TP_c\hat x} =\dfrac{1}{\sqrt{\hat x^TP_c\hat x}}\Big(-\hat x^T Q\hat x+2\hat x^TP\phi(\hat x,u)\Big)\crr{0}
&\quad+\dfrac{1}{\sqrt{\hat x^TP_c\hat x}}(2\hat x^TP \Big( BK(x-\bar{x})+LCz-LC(z-\bar{z})\Big) \cr}                           
$$
where $Q=-(A+BK)^TP+P(A+BK)$. In virtue of $(H_1)$, one can write 
$$
\eqalign{
&\frac{d}{dt} 2\sqrt{\hat x^TP_c\hat x}\le -\frac{\eta_c\|\hat x\|^2}{\sqrt{\hat x^TP_c\hat x}} \crr{0}
&\quad +\frac{1}{\sqrt{\hat x^T P_c \hat x}}\Bigg[\|P\|\|\hat x\|\big( \|BK\| \|(x-\bar{x})\|\crr{-4}
&\hskip3cm +\|LC\|\|z\|+\|LC\|\|(z-\bar{z})\| \big)\Bigg] \crr{0}
&\le \frac{-\eta_c}{\sqrt{\lambda_{\max}(P_c)}}\|\hat x\| \crr{2}
&\quad +\frac{\big( \|BK\| \|(x-\bar{x})\|+\|LC\|\|z\|+\|LC\|\|(z-\bar{z})\| \big)}{\sqrt{\lambda_{\min}(P_c)}} \cr}
$$
which verifies assumption $(A_1)$. Analogously, using the candidate ISS Lyapunov function $2\sqrt{V_o}$, one can prove that $(A_2)$ holds. Furthermore, it is trivial to show that $(H_1),(H_2)$ imply $(A_3),(A_4)$.
\end{proof}

                                                                \end{partO}
%
%
%
                                                                \begin{partP}

\medskip
Therefore, applying Lemma~\ref{lemma1} to the system \eqref{eqlip}, and using Theorem~\ref{theorem1}, to the event--triggered observer--based controller ensures asymptotic convergence to the origin.

\medskip
\begin{corollary}
If $(H_1),(H_2),(H_3),(H_4)$ are verified, the event--triggered control policy~\eqref{ctrlp}, \eqref{obsp} and the control $u=K\bar{\hat x}$ ensure the asymptotic stability of the closed--loop system \eqref{closed-loop nonlinear system}.
\end{corollary}

\medskip
\begin{proof} Lemma~\ref{lemma5} ensures that $(A_1),(A_2),(A_3),(A_4)$ are verified. Then one applies Theorem~\ref{theorem1} to the system \eqref{closed-loop  nonlinear system}.
\end{proof}

\section{Simulations} 

The proposed methodology will be applied to a robot with a flexible link, used as a benchmark example in several papers dealing with Lipschitz observers (see for instance \cite{Raghavan 1994}, \cite{Aboky 2002}, \cite{Pagilla 2004}). The dynamics are in the form~\eqref{eqlip}, with

\begin{equation}\nonumber
\eqalign{
\dot x&=Ax + \phi(x,u) +BK\bar{\hat{x}} \crr{0}
\dot{\hat{x}}&=A \hat x+ \phi(\hat x,u) +BK\bar{\hat{x}}+LC\bar{z} \crr{0}
y&=C x   \cr}                           
\end{equation}
where
$$
\begin{aligned}
A&=\pmatrix{0 & 1 & 0 & 0\crr{2} -48.6 & -1.25 & 48.6 & 0\crr{2} 0 & 0 & 0 & 1\crr{2} 19.5 & 0 & -19.5 &  0}, \ B=\Big(\matrix{0 & 21.6 & 0 & 0}\Big)^T \\
C&=\pmatrix{1  &   0   &  0 & 0\crr{2} 0 & 1 & 0 & 0}, \hskip15pt \phi=\Big(\matrix{0 & 0 & 0 & 3.3\sin x_3}\Big)^T\hskip-4pt.
\end{aligned}
$$
One considers the control $u=K\bar{\hat x}$, with 
$$
K=\Big(\matrix{7.8428 & 1.1212 & -4.3666 & 1.1243}\Big)
$$
and the observer~\eqref{closed-loop nonlinear system:1}, with
$$
L=\pmatrix{9.3334  &  1.0001 \crr{0} -48.7804 & 22.3665 \crr{0} -0.0524  & 3.3194 \crr{0} 19.4066   & -0.3167}.
$$
The closed--loop equations are in the form~\eqref{closed-loop  nonlinear system}. The simulations have been performed considering the initial states 
$$
x(0)=\Big(\matrix{1 & 1 & 1 & 1}\Big)^T,\quad \hat x(0)=\Big(\matrix{0 & 0 & 0 & 0}\Big)^T.
$$
The theoretical values obtained on the triggering policy can be used but are too restrictive, due to the over--approximation on the convergence rate of the nonlinear observer and on the triggering parameter estimations. Via simulations it is possible to better tune the triggering parameters. It is worth noting that there is an order of magnitude of 100 between the theoretical value and the practical ones. We compared the result of a system controlled using triggering policy
$$
\eqalign{
t_{k_i+1}&=\min_t \{t\ge t_{k_i}+0.01,\mid  |u_i(t)-u_i(t_{k_i})|>0.2|u_i(t)|\}  \crr{0}
t_{k_j+1}&=\min_t \{t\ge t_{k_j}+0.01,\mid |y_j(t)-y_j(t_{k_j})|>0.2|y_j(t)|\}\cr}
$$
with the case in which $t_{k_i+1}=t_{k_i}+0.05$. The simulations show that for $t\in[0,2]$ s the system and observer are closed to the equilibrium, while at $t=2$ s an impulse drives the system away from equilibrium. Then, for $t\in[2,15]$ s, the system is stabilized at the origin by the proposed observer--based controller.

\begin{figure}[h!]
\centerline{\includegraphics[width=9cm]{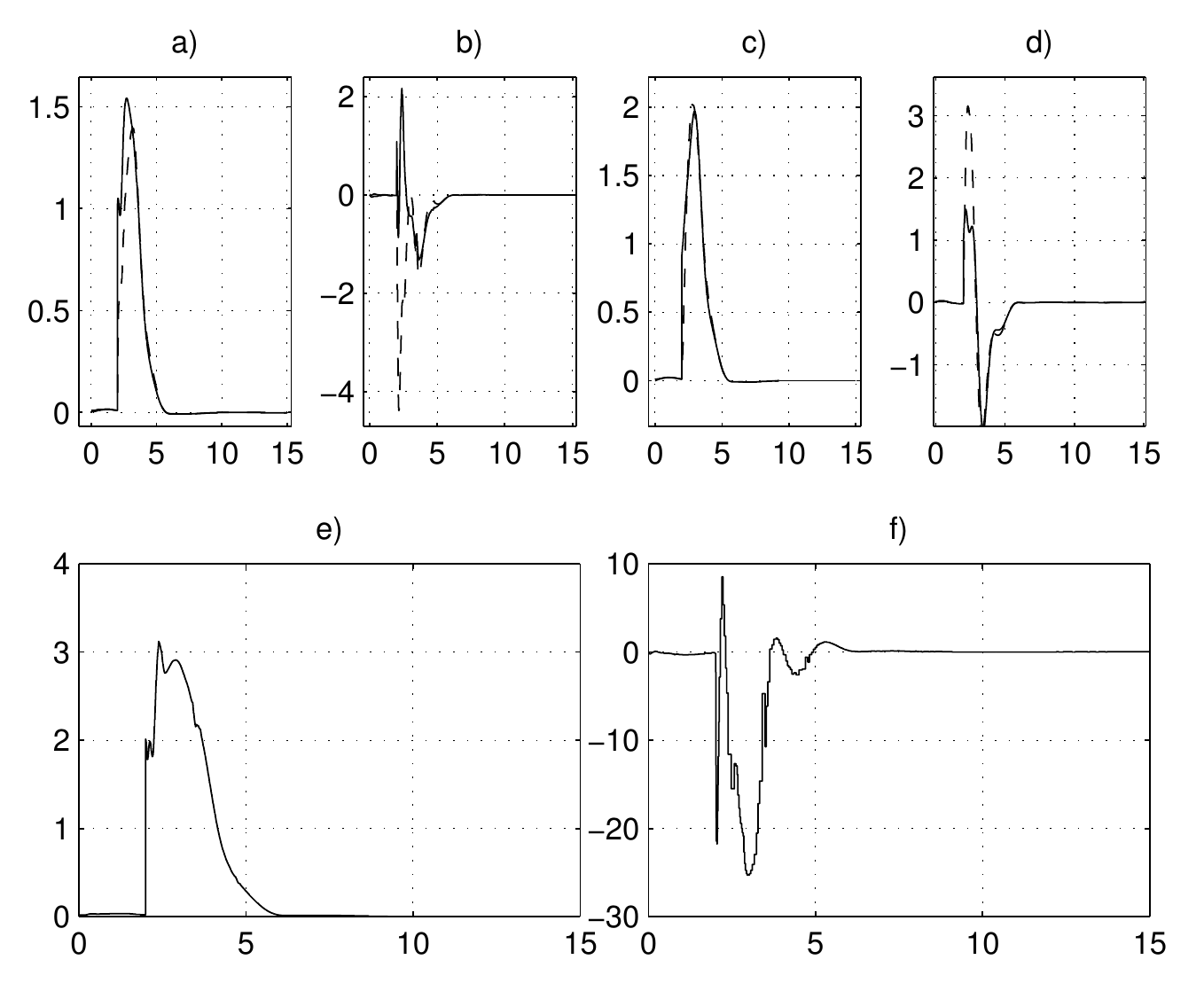}}
\vskip-.5cm\caption{System and observer state with the event--triggering: a) $x_1$, $\hat x_1$ ; b) $x_2$, $\hat x_2$; c) $x_3$, $\hat x_3$; d) $x_4$, $\hat x_4$; e) $\|x\|$; f) $u$.}\label{Figure 2}
%
\centerline{\includegraphics[width=9cm]{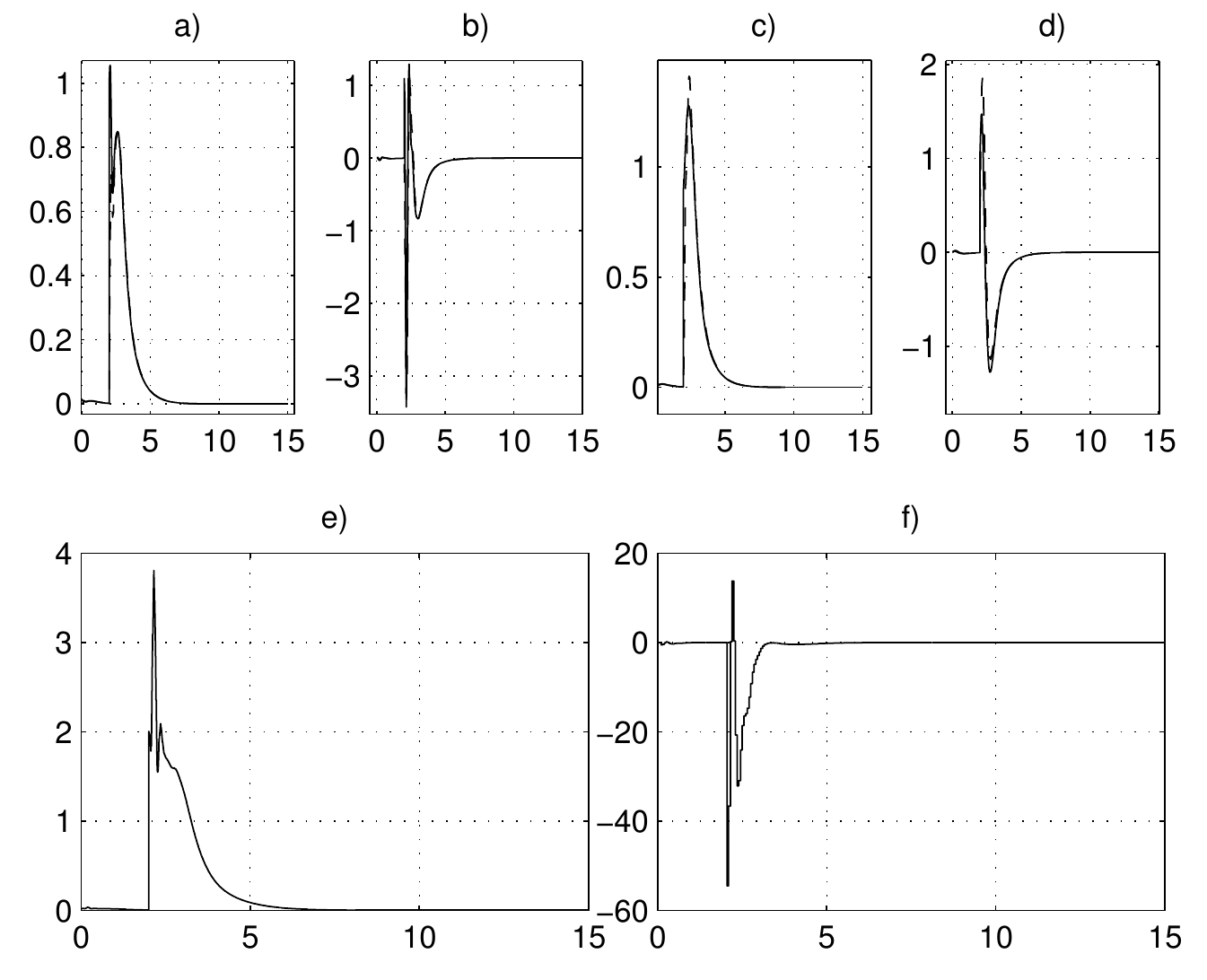}}
\vskip-.5cm\caption{System and observer state with periodic sampling: a) $x_1$, $\hat x_1$ ; b) $x_2$, $\hat x_2$; c) $x_3$, $\hat x_3$; d) $x_4$, $\hat x_4$; e) $\|x\|$; f) $u$.}
\label{Figure 3}
%
\centerline{\includegraphics[width=8.8cm]{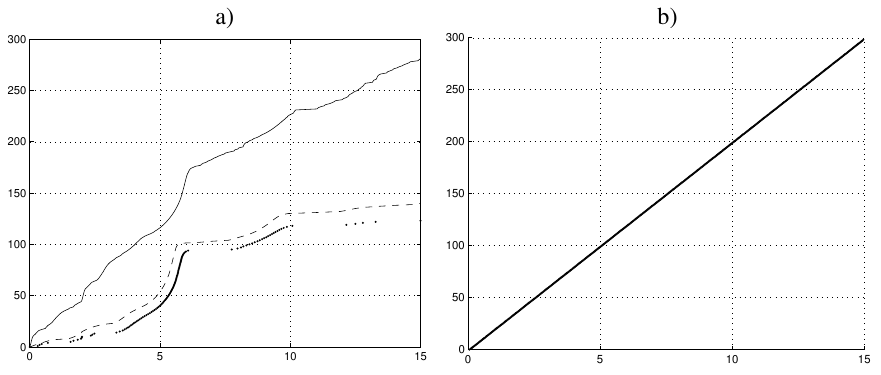}}
\caption{Number of triggers with the a)~Proposed event triggered policy; b)~Periodic sampling. $u_1$ (solid), $y_1$ (dashed), $y_2$ (dotted).}\label{Figure 4}
\end{figure}

Figs.~\ref{Figure 2}, \ref{Figure 3} show the convergence of the observer and the stabilization at the origin of the overall system. We can note that the event triggering is relatively slower with respect to the periodic sampling, but introduces a lower peaking. When confronting the number of triggers in Fig.s~\ref{Figure 4}.a, \ref{Figure 4}.b, it is clear that the number of communications is greater when considering the periodic sampling, so justifying the interest of the proposed event--triggering scheme. It is worth noting that the advantage of the method appears more clearly for output communications. As already noted, this is due to the fact that the observation and the control communications are done only when necessary. The comparison of Figs.~\ref{Figure 4}.a, \ref{Figure 4}.b illustrates a trade--off between ``intelligence'' in the sensor and actuator, and the communication burden.

\section{Conclusion}

In this paper we have presented an event--triggered observer--based controller for a class of nonlinear systems. Sufficient conditions in term of ISS stability for the observer and the observation error dynamics are given for designing an event--triggering mechanism ensuring the asymptotic convergence to the origin of the closed--loop system state. A particular subclass is that of systems with Lipschitz nonlinearities. The relevance of the approach has been highlighted by simulations of a robot with a flexible link, where the triggering parameters have been appropriately tuned.

Further work will include a practical way of determining theoretically a good choice of triggering parameters. Furthermore, even thought the hypotheses on the state and on the observer imply a separation principle (convergence of the observer without assumption on the trajectory of the state) when considering a continuous feedback, this property is lost when introducing the triggering policy. Since this is not the case when considering periodic sampling, an interesting question to address is: Can we ensure a separation principle when using event--triggered control policies?


                                                                \end{partP}
                                                                \begin{partQ}

\begin{enumerate}
\item The output feedback problem admits a certainty equivalence approach which requires a separation principle. I am sure that the authors are aware that this type of assumption is limited to a very special class of nonlinear systems. One should find an alternative motivation for the observer design problem considered here. 

\item Lemma 1 is not properly stated. I think that the authors require a global Lipschitz property for $b a_3^{-1}$ and $G$. Is this what is meant by Lipschitz on compacts?

LUCIEN: In the next version I removed the part taking Local Lipschitz hypothesis instead for ease of notation we deal with global Lipschitz function. And save local result for an ulterior version (where linearized system fall into the scope of example)

\item The main result Theorem goes along the line of existing results on event-triggered control systems. The contribution is to add a trigger for the measurement updates.  There is an underlying observability issue that seems to be missing here. Triggering on the process measurements will only highlight some states. It is quite possible that some of the state estimation errors do not vanish while the measurements do vanish.  

LUCIEN: the assumption A2 prevent this from happening since it require the capacity of synthesising an observer when continuous sampling is available. And the triggering mechanism will not produce singularity of observation.

\item The linear case study should probably clarify this situation but the result simply applies Theorem 1.

\item Typo in the proof of Lemma 1, $\| y_i - u_j\|$ should be $\|y_i - y_j\|$.

\medskip\hrule\medskip

\item One limitation of the proposed results is that no disturbance is present in system (1). It would be interesting to add bounded process and measurement noise.
This would make the analysis more significant: in this case, the maximum allowed inter-event time for a required asymptotic error would take into account the noise magnitude.

LUCIEN: Interesting but outside the scope of this article (Journal paper?)

\medskip\hrule\medskip

\item no comparison with other conventional methods has been done.

LUCIEN: The approach is novel in that it consider a new class of system and show that "classical" event trigger mechanism do work there is no point in comparing it with other triggering mechanism.

\item effect of measurement noise on results has not been considered.

LUCIEN: Another good point that would require further study

\item effect of initial condition on performance of proposed observer should be addressed.

LUCIEN: perhaps write a remark stating that A1 and A2 prevent harmful phenomenon to appear during the transient.

\item no consideration about control effort has been investigated.

LUCIEN: An important issue of this article is highlighted: We miss a good example.

\item effect of triggering parameters is very important in the proposed method and should be studied.

LUCIEN: The MAIN issue of the article is given that is our method to estimate valid parameter is in general restrictive

\end{enumerate}

                                                                \end{partQ}

                                                                \end{xcomment}
                                                                \end{document}